\documentclass[12pt]{article}
\usepackage{amssymb}

\usepackage{tikz}
\usetikzlibrary{arrows}

\newcommand{\be}{\begin{equation}}
\newcommand{\ee}{\end{equation}}
\def\bea{\begin{eqnarray}}
\def\eea{\end{eqnarray}}

\newcommand{\bn}{\begin{eqnarray}}
\newcommand{\en}{\end{eqnarray}}

\newcommand{\p}{\partial}

\newcommand{\nn}{\nonumber}

\newcommand{\no}{\noindent}

\def\bea{\begin{eqnarray}}
\def\eea{\end{eqnarray}}

\newcommand{\beq}{\begin{eqnarray}}
\newcommand{\eeq}{\end{eqnarray}}
\begin{document}

\title{\textbf{Parity singlets and doublets of massive spin-3 particles in $D=2+1$ via Nother gauge embedding }}
\author{D.Dalmazi$^{1}$\footnote{dalmazi@unesp.br}, E. L. Mendon\c ca$^{1}$\footnote{elias.leite@unesp.br}, A. L. R. dos Santos$^{2}$\footnote{alessandroribeiros@yahoo.com.br} \\
	\textit{{1- UNESP - Campus de Guaratinguet\'a - DFQ} }\\
	\textit{{CEP 12516-410, Guaratinguet\'a - SP - Brazil} }\\
	\textit{{2- Instituto Tecnol\'ogico de Aeron\'autica - DCTA} }\\
	\textit{{CEP 12228-900, S\~ao Jos\'e dos Campos - SP - Brazil} }\\}
\date{\today}
\maketitle

\begin{abstract}
Here we demonstrate  that the sixth order (in derivatives) spin-3 self-dual model can be obtained from the fifth order 
self-dual model via a Noether Gauge Embedding  (NGE) of longitudinal Weyl transformations
 $\eta_{(\mu\nu}\partial_{\alpha)}\Phi$. 
In the case of doublet models we can show that the massive spin-3 Singh-Hagen theory is dual to a fourth and to a sixth order theory, 
via a double round of the NGE procedure by imposing traceless longitudinal (reparametrization-like) symmetries $\partial_{(\mu}\tilde{\xi}_{\nu\alpha)}$ in the first round and transverse Weyl transformations $\eta_{(\mu\nu}\psi^T_{\alpha)}$  in the second one. 
Our procedure automatically furnishes the dual maps between the corresponding fields.
\end{abstract}
\newpage

\section{Introduction}

So far the observed elementary particles in nature have spin $s=0,1/2,1$. 
In principle, we could have elementary particles of arbitrary
integer or half-integer spin. In particular, in the spectrum of superstring theories, 
there are massive particles of arbitrarily high spin. As we increase
the spin we need higher rank tensors for a Lorentz covariant description. 
The higher the rank, the more redundant fields are introduced
which must be eliminated afterwards in order to achieve the correct counting of degrees
of freedom, namely, $2s+1$ in the massive case and $2$ for massless particles which corresponds to the
helicities $+s$ and $-s$. Some of the fields are called auxiliary fields. They have no physical content
but their equations of motion lead to
nontrivial constraints required for the reduction of degrees of freedom.

From the theoretical point of view, the main difficult in describing higher spin particles lies in the fact
that some of the auxiliary fields may stop being purely auxiliary due to interactions. They acquire a nontrivial
dynamics and we end up with
an incorrect number of degrees of freedom, some of them become ghosts. 

In $D=2+1$ dimensions it is possible to trade 
auxiliary fields into local symmetries by going to dual models of higher order in derivatives which acquire
more symmetries as the number of derivatives is increased. Since it is easier to control local symmetries (gauge symmetries)  than
the dynamics of auxiliary fields it is of interest to investigate this trading procedure. 
In particular, in $D=2+1$ we can define the so called self-dual models which are parity singlets of spin-s 
and of order $j$ in derivatives, henceforth $SD_j^{(s)}$. They describe 
massive particles of a given helicity $+s$ or $-s$ in a local way. By means of a Noether gauge embedding (NGE) procedure one
can go from $SD_j^{(s)}$ to $SD_{j+1}^{(s)}$. This has done in \cite{clovis}, \cite{mls} and \cite{SD_4} respectively
for spin $s=1,3/2$ and $s=2$. In all those cases $j$ runs from $1$ until the top value $2s$. In the spin-3 case,
further examined here, we have partially succeeded \cite{nge}  in going from $j=1$ until $j=4$ along
the $NGE$ approach.  Here we show in section 2 how to go from the model $SD_5^{(3)}$ of \cite{ddhigher}
to the top model $SD_6^{(3)}$ of \cite{bhth}. We still have a gap between $SD_4^{(3)}$ and $SD_5^{(3)}$.

Moreover, the $NGE$ also works for parity doublets containing both helicities $+s$ and $-s$. In \cite{SD_4}
we have obtained the fourth order linearized ``New Massive Gravity'' of \cite{bht} from the usual
second order Fierz-Pauli (FP) \cite{fp}
theory which describes massive spin-2 particles.  Here we derive in section 3 a fourth and sixth order spin-3 doublet model from
the second order spin-3 Singh-Hagen model \cite{sh}. We conjecture that there is chain of parity doublet models
of order $2,4,6,\cdots, 2s$ for arbitrary spin-s. 

\section{Higher derivative singlet models}

Here the spin-3 field is described in terms of a totally symmetric rank-3 tensor $h_{\mu\nu\alpha}$. There are some ``geometrical'' objects similar to those we know from general relativity like the Einstein and Schouten tensors which are given by:
\bea
\mathbb{G}_{\mu\nu\alpha}=\mathbb{R}_{\mu\nu\alpha}-\frac{1}{2}\eta_{(\mu\nu}\mathbb{R}_{\alpha)}\quad,\quad\mathbb{S}_{\mu\nu\alpha}=\mathbb{R}_{\mu\nu\alpha}-\frac{1}{8}\eta_{(\mu\nu}\mathbb{R}_{\alpha)}\eea
\no where the spin-3 Ricci tensor and its vector contraction have been introduced in \cite{deserdamour}, namely:
\bea \mathbb{R}_{\mu\nu\alpha}&=&\square{h}_{\mu\nu\alpha}-\partial^{\beta}\partial_{(\mu}h_{\beta\nu\alpha)}+\partial_{(\mu}\partial_{\nu}h_{\alpha)}\\
\mathbb{R}_{\alpha}&=&\eta^{\mu\nu}\mathbb{R}_{\mu\nu\alpha}=2\square{h}_{\alpha}-2\partial^{\beta}\partial^{\lambda}h_{\beta\lambda\alpha}+\partial_{\alpha}\partial^{\beta}h_{\beta}.\eea 

\no Along this work we use the mostly plus metric $(-,+,+)$ and unnormalized symmetrization: $(\alpha\beta\gamma)=\alpha\beta\gamma+\beta\gamma\alpha+\gamma\alpha\beta$. It is also often the use of the anti-symmetric operator $E_{\mu\nu}=\epsilon_{\mu\nu\alpha}\p^{\alpha}$ where $(Eh)_{\mu\nu\alpha}\equiv(2/3)E_{(\mu}^{\;\;\;\beta}h_{\beta\nu\alpha)}$. Given another totally symmetric tensor $j_{\mu\nu\alpha}$, the operators $\mathbb{G}_{\mu\nu\alpha}$ and $\mathbb{S}_{\mu\nu\alpha}$ are hermitian in the sense that:

	\be \mathbb{G}_{\mu\nu\alpha}[\mathbb{S}(h)]j^{\mu\nu\alpha}=\mathbb{S}_{\mu\nu\alpha}(h)\mathbb{G}^{\mu\nu\alpha}(j)=\mathbb{S}_{\mu\nu\alpha}(j)\mathbb{G}^{\mu\nu\alpha}(h)=h_{\mu\nu\alpha}\mathbb{G}^{\mu\nu\alpha}[\mathbb{S}(j)]\ee

A great advantage of the higher order self-dual models introduced in \cite{ddhigher}, is the absence of auxiliary fields, this is a key issue when we add interactions since auxiliary fields may become dynamic and destroy the correct counting of degrees of freedom. Here we revisit the equivalence of those models under the point of view of the $NGE$ approach, which reveals the role of the symmetries. The fifth order self-dual model ($SD_5^{(3)}$) for the massive spin-3 particle is given by:
\bea S_{SD_5^{(3)}}=\int{d^{3}x}\Big[-\frac{1}{2m^{2}}\mathbb{S}_{\mu\nu\alpha}(h)\mathbb{G}^{\mu\nu\alpha}(h)+\frac{1}{4m^{3}}\mathbb{S}_{\mu\nu\alpha}(h)\mathbb{G}^{\mu\nu\alpha}(E\,h)\Big]\label{SD_5}\eea

\no The whole action $S_{SD_5^{(3)}}$ is invariant under the gauge transformation
\be \delta_{\tilde{\xi},\psi^{T}}{h}_{\mu\nu\alpha}=\partial_{(\mu}\tilde{\xi}_{\nu\alpha)}+\eta_{(\mu\nu}\psi^{T}_{\alpha)}\label{Tr-4},\ee
where the parameter $\tilde{\xi}_{\nu\alpha}$ is symmetric and traceless while the the vector parameter $\psi_{\alpha}^T$ of the Weyl transformation is transverse ($\p^{\alpha}\psi_{\alpha}^T=0$). Besides, the fifth order term has an additional symmetry,
\be \delta_{\xi,\psi}{h}_{\mu\nu\alpha}=\partial_{(\mu}\tilde{\xi}_{\nu\alpha)}+\eta_{(\mu\nu}\psi^T_{\alpha)}+ \eta_{(\mu\nu}\p_{\alpha)} \phi=\partial_{(\mu}{\xi}_{\nu\alpha)}+\eta_{(\mu\nu}\psi_{\alpha)} \label{Tr-5}\ee

\no where the parameter $\xi_{\nu\alpha}$ is symmetric while $\psi_{\alpha}$ is an ordinary vector.

Once the additional symmetry, the longitudinal Weyl transformation, of the fifth order term is broken by the fourth order one, we would like to impose such symmetry to the model (\ref{SD_5}) in order to obtain a sixth order model ($SD_6^{(3)}$), which is invariant under (\ref{Tr-5}) but with the same particle content of the $SD_5^{(3)}$. We begin by adding a source term $j^{\mu\nu\alpha}$ coupled to a totally symmetric dual field $\tau^{\ast}_{\mu\nu\alpha}$:
\bea S_{SD_5^{(3)}}[j]=\int{d^{3}x}\Big[-\frac{1}{2m^{2}}\mathbb{S}_{\mu\nu\alpha}(h)\mathbb{G}^{\mu\nu\alpha}(h)+\frac{1}{4m^{3}}\mathbb{S}_{\mu\nu\alpha}(h)\mathbb{G}^{\mu\nu\alpha}(E\,h)+\tau^{\ast}_{\mu\nu\alpha}j^{\mu\nu\alpha}\Big]\label{sd5j}\eea
Notice that the dual field is chosen in such a way that it preserves the gauge invariance under (\ref{Tr-4}), then we have the fourth order dual field: $\tau^{\ast}_{\mu\nu\alpha}=(1/m^{4})\mathbb{G}_{\mu\nu\alpha}(\mathbb{S}(h))$. From (\ref{sd5j}) we now take the Euler tensor:

\bea K^{\mu\nu\alpha} \equiv \frac{\delta S_{SD_5^{(3)}}}{\delta h_{\mu\nu\alpha}}=-\frac{1}{m^{2}}\mathbb{G}^{\mu\nu\alpha}[\mathbb{S}(h)]+\frac{1}{2m^{3}}\mathbb{G}^{\mu\nu\alpha}[\mathbb{S}(Eh)]+\frac{1}{m^{4}}\mathbb{G}^{\mu\nu\alpha}[\mathbb{S}(j)],\label{Euler-SD_5}\eea

\no in order to implement a first iteration which is given by
\bea S_{1}=S_{SD_5^{(3)}}+\int{d^{3}x}\,\,K^{\mu\nu\alpha}a_{\mu\nu\alpha}\eea

\no where $a_{\mu\nu\alpha}$ is an auxiliary field. By taking the gauge variation of $S_1$ with respect to (\ref{Tr-5}) and choosing $\delta{a}_{\mu\nu\alpha}=-\delta{h}_{\mu\nu\alpha}$, we obtain:

\be \delta_{\xi,\psi}S_{1}=\int{d^{3}x}\;a_{\mu\nu\alpha}\delta_{\xi,\psi}{K}^{\mu\nu\alpha}.\ee

\no By calculating the variation of the Euler tensor we have then



\bea \delta_{\xi,\psi}{S}_{1}=\int{d^{3}x}\;\delta_{\xi,\psi}\left[\frac{1}{2m^{2}}\mathbb{S}_{\mu\nu\alpha}(a)\mathbb{G}^{\mu\nu\alpha}(a)\right],\eea

\no which automatically takes us to the second iteration, which is gauge invariant by construction and given by:

\bea S_{2}&=&S_{1}-\int{d^{3}x}\frac{m}{2}\mathbb{S}_{\mu\nu\alpha}(a)\mathbb{G}^{\mu\nu\alpha}(a)\nn\\
&=&\int{d^{3}x}\Big[{\mathcal{L}}_{SD_5^{(3)}}(h)-\frac{1}{m^{2}}\mathbb{S}_{\mu\nu\alpha}(a)\mathbb{G}^{\mu\nu\alpha}(h)+\frac{1}{2m^{3}}\mathbb{S}_{\mu\nu\alpha}(a)\mathbb{G}^{\mu\nu\alpha}(Eh)-\frac{1}{2m^{2}}\mathbb{S}_{\mu\nu\alpha}(a)\mathbb{G}^{\mu\nu\alpha}(a)\nn\\
&&\qquad+\frac{1}{m^{4}}\mathbb{S}_{\mu\nu\alpha}(a)\mathbb{G}^{\mu\nu\alpha}(j)+\tau^{\ast}_{\mu\nu\alpha}j^{\mu\nu\alpha}\Big].\eea

\no Integrating over the auxiliary fields $a_{\mu\nu\alpha}$ we have:

\bea S_{2}&=&\int{d^{3}x}\Bigg[-\frac{1}{4m^{3}}\mathbb{S}_{\mu\nu\alpha}(h)\mathbb{G}^{\mu\nu\alpha}(Eh)+\frac{1}{8m^{4}}\mathbb{S}_{\mu\nu\alpha}(Eh)\mathbb{G}^{\mu\nu\alpha}(Eh)\nn\\
&&\qquad\quad+\frac{1}{2m^{5}}\mathbb{S}_{\mu\nu\alpha}(Eh)\mathbb{G}^{\mu\nu\alpha}(j)+\frac{1}{2m^{6}}\mathbb{S}_{\mu\nu\alpha}(j)\mathbb{G}^{\mu\nu\alpha}(j)\nn\\
&&\qquad\quad-\frac{1}{2m^{2}}\mathbb{S}_{\mu\nu\alpha}\Big(a+h-\frac{Eh}{2m}-\frac{j}{m^{2}}\Big)\mathbb{G}^{\mu\nu\alpha}\Big(a+h-\frac{Eh}{2m}-\frac{j}{m^{2}}\Big)\Bigg].\eea

\no Notice that by shifting the auxiliary fields in such a way that $a_{\mu\nu\alpha}\rightarrow{a}_{\mu\nu\alpha}-h_{\mu\nu\alpha}+(1/2m)(Eh)_{\mu\nu\alpha}+(1/m^{2})j_{\mu\nu\alpha}$ we get a completely decoupled term depending on $a_{\mu\nu\alpha}$ which is free of particle content, see \cite{deserdamour}, and will be neglected henceforth. This allow us to obtain the sixth order self-dual model \footnote{We have used the following properties: $\mathbb{S}_{\mu\nu\alpha}(Eh)\mathbb{G}^{\mu\nu\alpha}(Eh)=\mathbb{S}_{\mu\nu\alpha}(h)\mathbb{G}^{\mu\nu\alpha}(E^{2}h)$.}:

\bea S_{SD_6^{(3)}}=\int{d^{3}x}\Big[-\frac{1}{4m^{3}}\mathbb{S}_{\mu\nu\alpha}(h)\mathbb{G}^{\mu\nu\alpha}(Eh)+\frac{1}{8m^{4}}\mathbb{S}_{\mu\nu\alpha}(h)\mathbb{G}^{\mu\nu\alpha}(E^{2}h)+h^{\ast}_{\mu\nu\alpha}j^{\mu\nu\alpha}+\mathcal{O}(j^{2})\Big]\label{SD_6}\eea

\no with the fifth order dual field
\be h^{\ast}_{\mu\nu\alpha}\equiv\frac{1}{2m^{5}}\mathbb{G}_{\mu\nu\alpha}[\mathbb{S}(Eh)].\ee

The sixth order self-dual model obtained here, is precisely the one first found in \cite{bhth} and investigated by some of us  in \cite{ddhigher}. It is invariant under a large set of gauge symmetries in the sense that  $\tilde{\xi}_{\mu\nu}\to \xi_{\mu\nu}$ and $\psi_{\alpha}^T\to \psi_{\alpha}$. Once again we stress that such self-dual descriptions do not need auxiliary fields, differently of the doublet models we are going to address in the next section. 

Finally, we can verify the classical equivalence between the $SD_5^{(3)}$ and the $SD_6^{(3)}$ models at the level of the equations of motion. From (\ref{SD_5}), we have:
\bea -\frac{1}{m^{2}}\mathbb{G}_{\mu\nu\alpha}(\mathbb{S}(h))+\frac{1}{6m^{3}}E_{(\mu}^{\;\;\;\beta}\mathbb{G}_{\beta\nu\alpha)}(\mathbb{S}(h))=0\; ;\eea

\no which in terms of the dual fields $\tau^{\ast}_{\mu\nu\alpha}$ give us:
\bea -m^2\tau^{\ast}_{\mu\nu\alpha}+\frac{m}{6}E_{(\mu}^{\;\;\;\beta}\tau^{\ast}_{\beta\nu\alpha)}=0\label{eom-SD_5}\;.\eea

In the other hand, the equations of motion from (\ref{SD_6}) with $j^{\mu\nu\alpha}=0$ are given by:
\bea -\frac{1}{2m^{3}}\mathbb{G}_{\mu\nu\alpha}(\mathbb{S}(Eh))+\frac{1}{12m^{4}}E_{(\mu}^{\;\;\;\beta}\mathbb{G}_{\beta\nu\alpha)}(\mathbb{S}(Eh))=0\;.\eea

\no Again, rewritten it in terms of the dual field $h^{\ast}_{\beta\nu\alpha}$ we have:

\bea -m^2h^{\ast}_{\mu\nu\alpha}+\frac{m}{6}E_{(\mu}^{\;\;\;\beta}h^{\ast}_{\beta\nu\alpha)}=0\label{eom-SD_6}\;.\eea

\no Then, we have showed that the $SD_5^{(3)}$ equations of motion (\ref{eom-SD_5}) can be taken to the $SD_6^{(3)}$ equations through the dual map $\tau^*_{\beta\nu\alpha} \to h^*_{\beta\nu\alpha}$ once they have the same form.

\section{Higher derivative doublet models}

Here we complement some previous discussions that we have made in \cite{hds3} where we have suggested master actions interpolating among three equivalent doublet models describing massive spin $3$ particles in $D=2+1$ dimensions. We have verified that the Singh-Hagen model is in fact dual to a fourth order model, which is analogue to the spin-2 New Massive Gravity model \cite{bht} . However in the spin-3 case differently of the spin-2 case one can obtain a sixth order model which has no analogue in the spin-2 context. After revisiting this issue under the point of view of symmetries some other analogies arise. In order to understand the role of symmetries when we are mapping such dual descriptions  we start with the massive second order Singh-Hagen model. The model requires a totally symmetric field $\phi_{\mu\nu\alpha}$ and auxiliary fields which may be either a vector or a scalar field.  Here to keep the similarities with our previous work, we choose scalar fields $W$:

\bea S_{SH} &=& \int d^3x \left[\frac{1}{2}\phi_{\mu\nu\lambda} {\mathbb G}_{\mu\nu\lambda}(\phi)-\frac{m^{2}}{2}(\phi_{\mu\nu\lambda}\phi^{\mu\nu\lambda}-3\phi_{\mu}\phi^{\mu}) -m\, \phi_{\mu}\p^{\mu}W\right]+S_1[W].\label{master}\eea

 \no The auxiliary action $S_1[W]$, is given by:

\be S_1[W]=\int d^3x \left(9m^2W^{2}-\frac{4}{3}W\Box W\right).\ee

\no With respect to the symmetries one can easily verify that the second order rank-3 term is invariant under traceless reparametrizations  $\delta_{\tilde{\xi}}\phi_{\mu\nu\lambda}=\p_{(\mu}\tilde{\xi}_{\nu\lambda)}$ . From the equations of motion with respect the rank-3 field, we have the Euler tensor:

\be
K^{\mu\nu\lambda}={\mathbb G}^{\mu\nu\lambda}(\psi)-m^2(\phi^{\mu\nu\lambda}-\eta^{(\mu\nu}\phi^{\lambda)})-\frac{m}{3}\eta^{(\mu\nu}\p^{\lambda)}W.\label{M}\ee

\no It is also convenient to keep in hand the trace of (\ref{M}) which is given by:

\be K^{\lambda}=\mathbb{G}^{\lambda}(\phi)+4m^2\phi^{\lambda}-\frac{5m}{3}\p^{\mu}W.\ee 

Introducing an extra auxiliary field $a_{\mu\nu\lambda}$ with the specific gauge symmetry $\delta_{\tilde{\xi}}a_{\mu\nu\lambda}=-\delta_{\tilde{\xi}}\phi_{\mu\nu\lambda}$ we have the first iteration:
\be S_1=S_{SH}+\int d^3x\,\, a_{\mu\nu\lambda}K^{\mu\nu\lambda}.\label{fi}\ee
In (\ref{fi}) we now perform the $\tilde{\xi}$-gauge variation wich after some calculation take us to the following result:

\be\delta_{\tilde{\xi}} S_1=\int \,\,d^3x\, \delta_{\tilde{\xi}}\left\lbrack \frac{m^2}{2}(a_{\mu\nu\lambda}a^{\mu\nu\lambda}-3a_{\mu}a^{\mu})\right\rbrack,\ee

\no which by construction allows us to determine the second iterate action automatically $\tilde{\xi}$-gauge invariant given by: 

\be S_2= S_{SH}+\int\,\,d^3x\,\left\lbrack a_{\mu\nu\lambda}K^{\mu\nu\lambda}-\frac{m^2}{2}(a_{\mu\nu\lambda}a^{\mu\nu\lambda}-3a_{\mu}a^{\mu})\right\rbrack\ee

\no solving the equations of motion for the auxiliary fields $a_{\mu\nu\lambda}$, one can invert it in terms of the Euler tensors, which then give us:

\be S_2=S_{SH}+\frac{1}{2m^2}\int\,\,d^3x\,\left(K_{\mu\nu\lambda}K^{\mu\nu\lambda}-\frac{3}{4}K_{\mu}K^{mu}\right),\label{S2}\ee

\no by substituting back the Euler tensor in the expression (\ref{S2}), we finally have the fourth order model:
\bea S_4&=&\int d^3x \, \left[-\frac{1}{2}\phi_{\mu\nu\lambda}{\mathbb G}^{\mu\nu\lambda}(\phi)+\frac{1}{2m^2}
{\mathbb S}_{\mu\nu\lambda}(\phi){\mathbb G}^{\mu\nu\lambda}(\phi)+\frac{1}{12m}\phi_{\mu\nu\lambda}{\mathbb G}^{\mu\nu\lambda}(\eta\p W)\right]+S_2[W].\nn\\
\label{fourth}\eea

\no Where $\eta\p W$ stands for the fully symmetric tensor  $\eta_{(\mu\nu}\p_{\rho)}W = \eta_{\mu\nu}\p_{\rho}W
+ \eta_{\nu\rho}\p_{\mu} W + \eta_{\rho\mu} \p_{\nu} W $, while

\be S_2[W]=\int d^3x \left(9m^2W^{2}-\frac{9}{8}W\Box W\right). \label{s2w}\ee

\no The fourth order model that we have obtained in (\ref{fourth}) is precisely the one we have found in \cite{hds3}. There we have also added source terms in order to verify the dual map with the equations of motion of the Singh-Haggen model. One also notices that the auxiliary action as well as the linking term between $\phi_{\mu\nu\alpha}$ and the auxiliary fields $W$, have been automatically corrected during the process, which is a fundamental step in order to get rid of the lower spin propagation modes, which in this case is a spin-0 mode.

\section{From the fourth  to the sixth order model }
The action (\ref{fourth}) is invariant under the traceless reparametrization $\delta_{\tilde{\xi}}\phi_{\mu\nu\lambda}$, but once the fourth order term is indeed the same one we have in the fifth order self-dual model (\ref{SD_5}), we know that it is invariant under an additional gauge symmetry given by transverse Weyl transformation $\delta_{\psi^T}\phi_{\mu\nu\lambda}=\eta_{(\mu\nu}\psi_{\lambda)}^T$. Such symmetry is broken by the first term of the Singh-Hagen action, which indicates that there is another round of $NGE$ in order. To implement this symmetry we start by calculating the Euler tensor from (\ref{fourth})  which is given by:

\be
K^{\mu\nu\lambda}=-{\mathbb {G}}^{\mu\nu\lambda}(\phi)+\frac{1}{m^2}{\mathbb {G}}^{\mu\nu\lambda}\left[S(\phi)\right]+\frac{1}{12m}{\mathbb {G}}^{\mu\nu\lambda}(\eta\p W)\label{eu}.\ee

\no Again, an auxiliary field is suggested in a first iterated action:

\be S_1=S_{SH}-\int d^3x\,\, a_{\mu\nu\lambda}K^{\mu\nu\lambda}.\label{fi2}\ee

When we take the $\psi_{\lambda}^T$ gauge-transformation on $S_1$ we end up with the following result, after some calculation:

\be \delta_{\psi^T}S_1=\int \,\,d^3x\,\delta_{\psi^T}\left[\frac{a_{\mu\nu\lambda}\mathbb{G}^{\mu\nu\lambda}(a)}{2}\right].\ee

\no As we have seen before, now we have a gauge invariant action given by:

\be S_2=S_4-\int\,d^3x\,\,\left[a_{\mu\nu\lambda}K^{\mu\nu\lambda}+\frac{a_{\mu\nu\lambda}\mathbb{G}^{\mu\nu\lambda}(a)}{2}\right]\ee

\no One can notice that the Euler tensor given at (\ref{eu}) can be rewritten in such a way that $K_{\mu\nu\lambda}=\mathbb{G}^{\mu\nu\lambda}(b)$ where 

\be b^{\mu\nu\lambda}=-\phi^{\mu\nu\lambda}+\frac{1}{m^2}\mathbb{S}^{\mu\nu\lambda}(\phi)+\frac{1}{12m}\eta^{(\mu\nu}\p^{\lambda)}W\ee

\no which allows us to rewrite the action $S_2$ as:

\be S_2=S_4-\int\,d^3x\,\, \left[\frac{1}{2}(a_{\mu\nu\lambda}+b_{\mu\nu\lambda})\mathbb{G}^{\mu\nu\lambda}(a+b)-\frac{1}{2}b_{\mu\nu\lambda}\mathbb{G}^{\mu\nu\lambda}(b)\right]\label{s6d}\ee

\no shifting the auxiliary field $a_{\mu\nu\lambda}\to a_{\mu\nu\lambda}-b_{\mu\nu\lambda}$ we get a completely decoupled second order term, which is free of particle content, see \cite{deserdamour}. After substituting back $b_{\mu\nu\lambda}$ in (\ref{s6d}) we have after some rearrangements  a sixth order action invariant under the gauge transformations (\ref{Tr-4}). 

\bea S_6&=&\int d^3x\,\left[- \frac{1}{2m^2}{\mathbb S}_{\mu\nu\lambda}(\phi){\mathbb G}^{\mu\nu\lambda}(\phi)+\frac{1}{2m^4}{\mathbb S}_{\mu\nu\lambda}(\phi){\mathbb G}^{\mu\nu\lambda}[{\mathbb S}(\phi)]+\frac{1}{12m^3}{\mathbb S}_{\mu\nu\lambda}(\phi){\mathbb G}^{\mu\nu\lambda}(\eta\p W)\right].\nn\\
&+& S_3[W]\label{HD}\eea

\no Notice that the auxiliary action $S_3[W]$ has now an extra higher derivative term :

\be S_3[W]=\int d^3x \left(9m^2W^{2}-\frac{9}{8}W\Box W+\frac{9}{64m^2}W\Box^2 W\right).\ee

The sixth order spin-3 model \cite{hds3} and the fifth order self-dual model $SD_5^{(3)}$ (\ref{SD_5}) share the same symmetries (\ref{Tr-4}). This is similar to the spin-2 case where the $NMG$ ($4th$ order) and the Topologically Massive Gravity $TMG$ \cite {DJT}, ($3rd$ self-dual model $SD_3^{(2)}$) have the same symmetries.

\section{Conclusion}

In the works \cite{clovis}, \cite{mls}, and \cite{SD_4} one has shown respectively, that
the spin-1, spin-3/2 and spin-2 self-dual models of j-th order in derivatives can be obtained 
from the models of previous (j-1)-th order via a Noether gauge embedding (NGE)  
procedure, where j runs from $2$ until the top value $2s$.  

Regarding the spin-3 case we have shown in \cite{nge} that such procedure only works until
the fourth order, i.e., j$=2,3,4$. Explicitly, the models and the 
symmetries\footnote{The field $\omega_{\mu (\alpha\beta)}$ satisfies $\omega_{\mu (\alpha\beta)}=\omega_{\mu (\beta\alpha)}$ and 
$\eta^{\alpha\beta}\omega_{\mu (\alpha\beta)}=0$.} used in the NGE procedure are sketeched below

\begin{center}
	\begin{tikzpicture}[>=latex',node distance = 5.25cm]
	\node (S1) {$SD_1^{(3)}$};
	\node [right of = S1] (S2) {$SD_2^{(3)}$};
	\node [right of = S2] (S3) {$SD_3^{(3)}$};
	\node [right of = S3] (S4) {$SD_4^{(3)}$};

	\draw [->,thick] (S1) to [ left=50] node[below] {$\delta\omega_{\mu(\nu\alpha)}=\partial_{\mu}\tilde{\xi}_{\nu\alpha}$} (S2);
	
	\draw [->,thick] (S2) to [ right=50] node[below] {$ \delta\omega_{\mu(\beta\gamma)}=\epsilon_{\mu\beta\rho}\Phi^{\rho}_{\,\,\gamma}+\epsilon_{\mu\gamma\rho}\Phi^{\rho}_{\,\,\beta}$} (S3);
	
	\draw [->,thick] (S3) to [ left=50] node[below] {$\delta\phi_{\mu\beta\gamma}=\partial_{(\mu}\xi_{\beta\gamma)}$} (S4);	
	\end{tikzpicture}
	
	\end{center}
	
\no In particular, we had not
been able to derive any fifth order spin-3 model via NGE.
Consequently, the top 6th-order spin-3 model $SD_6^{(3)}$   of \cite{bhth} 
could not be reached from the fourth order model $SD_4^{(3)}$
of \cite{nge}. 

Usually, a self-dual model of order j contains a j-th and a (j-1)-th order term.
The j-th term has more symmetries in general as compared to the rest of the Lagrangian. 
The exceeding symmetry is the one we use in the NGE approach.
It turns out that both fourth and third order terms inside  the $SD_4^{(3)}$ model
defined in \cite{nge}
are invariant under the same set of transformations ($\delta h_{\mu\nu\rho} = \p_{(\mu}\xi_{\nu\rho )}$) 
. So no difference is left over to be implemented in the NGE approach. Recently however,  we have 
found \cite{ddhigher}, by other means, the missing spin-3 fifth order self-dual model $SD_5^{(3)}$. Here we have 
shown that it is now possible to arrive at the $SD_6^{(3)}$ via  NGE of longitudinal Weyl transformations which is the symmetry of
the fifth order term of $SD_5^{(3)}$, not present in the fourth order term, namely

\begin{center}

\begin{tikzpicture}[>=latex',node distance = 7.0cm]
\node (S1) {$SD_5^{(3)}$};
\node [right of = S1] (S2) {$SD_6^{(3)}$};

\draw [->,thick] (S1) to [ left=50] node[below] 
{$\delta\phi_{\mu\beta\gamma}=\eta_{(\mu\nu}\partial_{\alpha)}\Phi$} (S2);

\end{tikzpicture}

\end{center}

\no We still do not know how to fill up  the gap between $SD_4^{(3)}$ and $SD_5^{(3)}$.
We believe that there might be another fourth order self-dual model whose embedding would lead us to $SD_5^{(3)}$.
Unfortunately we do no know how to go downstairs in derivatives systematically. This is still under investigation.

The NGE procedure also works for parity doublets, we
have shown in \cite{SD_4} that the fourth order spin-2  ``New Massive Gravity'' 
of \cite{bht}, in its linearized form, can be derived from the usual (second order)
Fierz-Pauli theory \cite{fp} via NGE of linearized reparametrizations $\delta h_{\mu\nu}=\p_{\mu}\xi_{\nu} + \p_{\nu}\xi_{\mu}$.
Here we have generalized \cite{SD_4} for spin-3 doublets. 
From the usual massive second order Singh-Hagen model we have derived a fourth and a sixth order dual doublet model
with helicities $+3$ and $-3$. Namely, 

\begin{center}
\begin{tikzpicture}[>=latex',node distance = 5.0cm]
\node (S1) {$S_{SH}$};
\node [right of = S1] (S2) {$S_4$};
\node [right of = S2] (S3) {$S_6$};

\draw [->,thick] (S1) to [ left=50] node[below] {$\delta\phi_{\mu\beta\gamma}=\partial_{(\mu}\tilde{\xi}_{\nu\lambda)}$} (S2);

\draw [->,thick] (S2) to [ right=50] node[below] {$\delta\phi_{\mu\beta\gamma}=\eta_{(\mu\nu}\psi_{\lambda)}^T$} (S3);

\end{tikzpicture}
\end{center}

\no We believe that there is a chain of $s$ dual doublet models of spin-s and order $j=2,4,6, \cdots ,2s$.
Differently of the spin-2 case, where the top fourth $(2s)$ order term (K-term) of the 
top doublet model (linearized NMG) coincides with the fourth order term of the 
top (4th order) spin-2 self-dual model, the sixth order term of the top doublet model $S_6^{(3)}$  does not coincide
with the sixth order term of the top singlet model $SD_6^{(3)}$. 
We are currently investigating the soldering
of two $SD_6^{(3)}$ models of opposite helicities in order to produce a doublet model without auxiliary fields, contrary to
$S_6$ which contains an auxiliary  scalar field. There is no doublet spin-3 model without auxiliary fields
even in $D=2+1$, to the best we know. 

If the soldering procedure can be successfully implemented we will be able to build up massive higher spin Lagrangians systematically
in $D=2+1$ and in $D=3+1$ (doublet models) since the doublet models have the same form in $D=2+1$ and in $D=3+1$.

\section{Acknowledgements}

The work of D.D. is partially supported by CNPq  (grant 306380/2017-0). A.L.R.dos S. is supported by a CNPq-PDJ
(grant 150524/2018-8).

\end{document}